\documentclass[11pt]{article}
\usepackage{bm,amsmath,color,dsfont,amssymb}
\usepackage{graphicx}
\usepackage[section]{placeins}
\usepackage[hidelinks]{hyperref}
\usepackage{caption}
\usepackage{cleveref}
\usepackage{booktabs}
\usepackage{tabularx}
\usepackage{tgtermes}
\usepackage{authblk}

\title{Non-Equilibrium Economics:\\ A Physicist's Point of View}
\author{Jean-Philippe Bouchaud\\
\small Capital Fund Management \& Acad\'emie des Sciences}
\date{}

\begin{document}
\maketitle

\begin{abstract}
Financial and economic history is strewn with bubbles and crashes, booms
and busts, crises and upheavals of all sorts. Understanding the origin of
these events is arguably one of the most important problems in economic
theory: are economies intrinsically unstable, and can one ``stabilize
unstable economies''? In this review I argue, from a physicist's vantage
point, that the concept of equilibrium --- so central to mainstream economic
thinking --- is likely to be the exception rather than the rule in large,
complex, interacting systems. Drawing on a series of stylized ``toy'' models,
I show how excess volatility, endogenous crises and crashes, inflation swells
and persistent inequalities can all emerge naturally from genuinely
out-of-equilibrium dynamics, without invoking large exogenous shocks. Three
generic mechanisms recur throughout: trapping in a multiplicity of
history-dependent equilibria; the impossibility of dynamically reaching
equilibrium, leading to oscillations and chaos; and the spontaneous evolution
towards fragile, marginally stable states --- the self-organized criticality
paradigm. I stress that these are phenomenological scenarios rather than
calibrated theories: there is, at this stage, no ``smoking gun''. But the
burden of proof, I contend, should be on the equilibrium camp.
\end{abstract}

%=====================================================================
\section{Introduction}
\label{sec:intro}
%=====================================================================

\paragraph{Excess volatility everywhere.} Financial and economic history is strewn with bubbles and crashes, booms and
busts, crises and upheavals of all sorts. Why? Are economies and financial
markets intrinsically stable, merely buffeted by a stream of external shocks,
or are they intrinsically \emph{unstable}, generating their own turbulence from
within? This question, forcefully posed long ago by Hyman Minsky
\cite{minsky,minsky2008stabilizing}, is arguably one of the most important ---
and most stubbornly unresolved --- problems in economic theory.

The empirical evidence for ``too much'' volatility is pervasive. The
year-on-year growth of the US GDP has been, since 1950, of the order of
$+3\% \pm 2.5\%$: the fluctuations are almost as large as the mean. Even in years where not much is happening in the world, large economies tend to fluctuate significantly. As John
Cochrane candidly put it, \emph{``What shocks are responsible for economic
fluctuations? Despite at least two hundred years in which economists have
observed fluctuations in economic activity, we still are not sure''}
\cite{cochrane1994shocks}. The financial-market analogue is the equally famous
``excess volatility puzzle'' elicited by Robert Shiller and by LeRoy and Porter
\cite{shiller1981stock,shiller1987volatility,leroy1981present}: asset prices
move far more than any reasonable measure of changing fundamentals would
warrant, and most of these swings occur without identifiable cause ---
\emph{``the evidence that large market moves occur on days without identifiable
major news casts doubts on the view that price movements are fully explicable by
news''} \cite{cutler1988moves,joulin2008,marcaccioli2022exogenous}. In
macroeconomics the corresponding conundrum is Bernanke's ``small shocks, large
business cycle'' puzzle \cite{Bernanke1996}. Standard equilibrium models --- in
particular Dynamic Stochastic General Equilibrium (DSGE) models, see e.g.
\cite{gali2015monetary} --- proved, by construction, ill-equipped to deal with
such endogenous jumps and crises, as was felt acutely in 2008. Olivier Blanchard
later conceded: \emph{``We in the field did think of the economy as roughly
`linear', constantly subject to different shocks, [\dots] but naturally
returning to equilibrium over time. [\dots] The main lesson of the crisis is
that we were much closer to `dark corners' --- situations in which the economy
could badly malfunction --- than we thought''} \cite{blanchard2014danger}.

These ``dark corners'' are, as I understand them, precisely the regions where
non-equilibrium effects and collective instabilities take over. My thesis is
thus based on a simple observation: classical economics relies heavily on the
concept of \emph{equilibrium}, but often says too little about the dynamics by
which equilibrium is reached --- or in fact not reached. This matters because the
existence of an equilibrium is not, by itself, enough to make it economically
relevant. What matters is whether decentralized learning, coordination, and
adjustment actually drive the system toward that equilibrium on empirically
reasonable time scales.

A useful way to sharpen the issue is to distinguish notions of equilibrium in
physics and in economics \cite{pangallo2024equations}.
In \emph{mechanical equilibrium}, nothing moves. In \emph{statistical
equilibrium}, everything moves, but the probability distribution is stationary,
i.e.\ the same today and tomorrow. In \emph{economic equilibrium}
\cite{gali2015monetary,acemoglu2012network}, firms maximize profits and
households maximize utility (with possible constraints or ``frictions''), and
markets clear given expectations, but equilibrium tomorrow might be very
different from equilibrium today. This framework bypasses the question of
adjustment: the standard shortcut is to assume that equilibrium effectively
re-establishes itself immediately after shocks.

\paragraph{Is equilibrium relevant to understand economic fluctuations?}
This leads to what I see as the central open problem: \emph{when is economic
equilibrium dynamically relevant?} The lesson from physics --- and the guiding
thread of this review --- is that in systems with many interacting agents, a
broad spectrum of time scales is the rule rather than the exception.
Metastability and hysteresis imply that the system may remain trapped for long
periods in local equilibria even when a more efficient state exists; more
radically, it may keep exploring local attractors without ever settling into a
globally preferred one. From this perspective, assuming fast equilibration is not
an innocuous simplification: it suppresses history, path-dependency and
coordination failures, even though these may be essential features of economic
reality.

Several major empirical puzzles can be read in these terms. In finance, excess
volatility remains difficult to reconcile with standard present-value logic; the
Efficient Market Hypothesis also sits uneasily alongside evidence that markets
digest order imbalances and portfolio shifts only gradually, as emphasized by
the literature on price impact and the ``inelastic markets'' hypothesis
\cite{bouchaud2009markets,toth2011anomalous,bouchaud2018trades, gabaix2021search}. In
macroeconomics, the ``small shocks, large business cycles'' puzzle suggests that
interactions, complementarities, and network propagation may matter more than
equilibrium models typically admit
\cite{acemoglu2012network,baqaee2019macroeconomic} --- although see
\cite{elliott2022networks}. More generally, feedback loops are everywhere: the
actions of some affect the beliefs of others, beliefs affect actions, and these
recursive effects can produce amplification, instability, and multiple
self-fulfilling outcomes, possibly generating large excess (non-fundamental)
volatility.

A second open problem concerns expectations and learning. One influential idea
is that rational expectations can emerge asymptotically through adaptive learning
\cite{evans2013learning,marcet1989convergence} --- an important attempt to
restore equilibrium foundations dynamically. But it raises difficult questions.
First, \emph{what is the relevant long run?} If convergence requires a
sufficiently stationary environment over a sufficiently long period, its
practical relevance may be limited in economies marked by structural change,
radical innovation, institutional evolution, and shifting policies; the issue is
not whether convergence is possible in principle, but whether it occurs on time
scales that matter. Second, \emph{can the rational equilibrium be reached at
all}, or are \emph{satisficing} equilibria the best one can ever learn in
complex situations \cite{simon1955behavioral,galla2013complex,PhysRevX.14.021039}?
In complex interacting environments, indeed, dynamics may fail in several
distinct ways --- converging only after astronomical times, becoming trapped in
suboptimal states, or never converging at all --- so that equilibrium
\emph{existence} and equilibrium \emph{attainability} become separate questions.
This is the theme of Sec.~\ref{sec:dynamics}, and it is visible in coordination
games, hysteresis phenomena, self-fulfilling prophecies, habit formation,
rewiring production networks, and self-referential strategic environments
\cite{hommes2021behavioral,dessertaine2022out,fosset2020endogenous,bouchaud2023self,colon}
--- the very examples around which this review is organized.

\paragraph{Putting dynamics at the centre stage.}
The methods best suited to these questions are those that put dynamics and endogenous fluctuations at the
center. Agent Based Models (ABM) are particularly promising, because they allow
one to study actual adjustment paths rather than comparative statics around a
fixed point, and accommodate heterogeneous, interacting, non-stationary
environments \cite{hommes2021behavioral,farmer2009economy,dosi2019more,farmer2024making, dosi2023foundations}.
Network models are equally important, since production, finance, and information
transmission all depend on interaction structures that may amplify shocks or
stabilize them \cite{acemoglu2012network,carvalho,dessertaine2022out}. More
generally, tools from complexity and statistical physics --- metastability,
hysteresis, spontaneous oscillations, self-organized criticality, and
quasi-nonergodicity --- seem well suited to economies in which aggregate
outcomes emerge from decentralized interaction across multiple time scales
\cite{arthur2015process,scheinkman1994self,bouchaud2013crises,gualdi2015endogenous,bouchaud2023self,Moran2024,moran2025critical}.
My own angle of attack, accordingly, is to build minimal, stylized models that
reproduce stylized facts \emph{generically} --- for broad swaths of parameter
space and without fine-tuning --- rather than to calibrate elaborate models that
forecast. The examples I discuss include firm networks with adaptive pricing,
production and rewiring rules \cite{dessertaine2022out,colon}; complex games with
self-referential beliefs \cite{PhysRevX.14.021039}; habit formation and
``keeping up with the Joneses'' mechanisms \cite{moran2020force,morelli2020confidence};
spontaneous synchronization phenomena \cite{gualdi2015endogenous}; and, more
generally, the possibility that macroeconomic fluctuations and crises are at
least partly endogenous rather than purely shock-driven
\cite{de2011animal,beaudry2015reviving,dessertaine2022out,martin2026resilient, portier202615}.

I want to stress at the outset that all the models discussed below are of this
phenomenological nature: they are meant to establish that excess volatility,
crises and inequalities \emph{can} arise endogenously out of equilibrium. They are
\emph{plausible generative scenarios}, not established truths; besides the
observation that economic and financial volatility is too high, there is at this
stage no smoking gun, no precise theoretical prediction from the non-equilibrium
models I will discuss that is unambiguously borne out by data (see however the
discussion, Sec.~\ref{sec:conclusion}).

\paragraph{Disequilibrium economics: an older tradition revisited.}
\label{sec:disequilibrium}
The viewpoint developed here is, in one sense, a return to an older tradition. As
Franklin Fisher stressed long ago in his neglected \emph{Disequilibrium
Foundations of Equilibrium Economics} \cite{fisher1989disequilibrium}, the very
question of \emph{how} --- and \emph{whether} --- equilibrium is reached is of
fundamental importance and cannot be swept under the rug. A basic conceptual
move that flourished in the 1960s and 1970s under the name of disequilibrium
macroeconomics was the following: when prices fail to clear markets in the short
run, agents face \emph{quantity constraints}~\cite{clower1965keynesian,leijonhufvud1968keynesian,barro1971general,benassy1975neo}.
In that framework the economy switches between \emph{regimes} depending on which
constraints bind --- Keynesian unemployment (both goods and labour markets in
excess supply), classical unemployment (goods in excess demand, labour in excess
supply), and repressed inflation (both in excess demand) --- and, crucially,
each regime calls for a different policy response: demand stimulus alleviates
Keynesian, but not classical, unemployment \cite{malinvaud1977theory}.

This program was abandoned not because it was empirically wrong
\cite{Backhouse_Boianovsky_2012}, but because the dynamics across regimes was
analytically intractable in any model with more than two markets, and because
the emphasis in macroeconomics moved, after the Lucas critique \cite{Lucas1976},
towards optimization-based microfoundations that stamped fixed-price assumptions
as ad hoc. The tradition has, however, seen a recent revival, see for example
\cite{Michaillat2015}. Many accounts, however, remain equilibrium-like in a
specific and important sense: the spillovers between rationed agents are resolved
\emph{instantaneously} --- the entire network is taken to have already settled
into a mutually consistent configuration between one instant and the next --- and
stability further requires that agents coordinate, from the outset, on a unique
forward-looking path. Both features presuppose a degree of instantaneous
coordination, with no representation of who observes which quantities, who acts
first, or how long a disruption takes to propagate from one agent to the next. It
is precisely here that the program embraced by ABMs and reviewed below parts
ways: the models that follow take the disequilibrium intuition seriously ---
quantity constraints, non-substitutable inputs, myopic adjustment rules --- but
insist on following the \emph{genuine dynamics} of propagation, coordination and
history, with no instantaneous fixed-point shortcut \cite{pangallo2024equations}.

My broader hypothesis is that many persistent puzzles in economics may not simply
reflect missing frictions around an otherwise adequate benchmark; they may
instead point to a deeper issue --- the benchmark itself is too static. If a
paradigm shift is on the horizon, it may come from taking disequilibrium
seriously, not as a temporary deviation from the main object of interest, but as
a central organizing principle in its own right.

\paragraph{Roadmap.}
After discussing when the ``equilibrium shortcut'' is legitimate and the three
generic ways in which dynamics can spoil it (Sec.~\ref{sec:dynamics}), I organize
the material around three recurring mechanisms, or ``flavours'', of
out-of-equilibrium behaviour. Section~\ref{sec:multiplicity} discusses
multiplicity of equilibria, mimetism and self-trapping; Sec.~\ref{sec:learning}
turns to learning in a complex world; Sec.~\ref{sec:firms} to coordination
breakdown in firm networks; and Sec.~\ref{sec:soc} to fragile optimization and
self-organized criticality. Section~\ref{sec:conclusion} discusses the (still
open) question of empirical discrimination and the policy consequences of
fragility.
%=====================================================================
\section{When Dynamics Spoils It All --- Lessons from Physics}
\label{sec:dynamics}
%=====================================================================

Having distinguished in the introduction the mechanical, statistical and
economic notions of equilibrium, let me now dwell on the feature that will
matter most in what follows: the crucial assumption, in the economic case, that
equilibrium is reached \emph{instantaneously}. Equilibrium is taken to adapt at
once to any (unanticipated) external shock; fluctuations of output are then, by
construction, nothing but differences between successive static equilibria
--- this is the so-called method of ``comparative statics''. As in
thermodynamics, this is a tremendously convenient assumption, precisely because
it allows one to by-pass entirely the description of the \emph{dynamics} towards
equilibrium, and of the \emph{history} of the system prior to its observation.
Successive equilibria, in economics, do not ``talk'' to each other. The question
raised in Sec.~\ref{sec:intro} --- \emph{when is equilibrium dynamically
relevant?} --- is exactly the question of whether this shortcut is legitimate.

Physics suggests it is often not, because equilibrium can take a very long time
to reach --- if it is reached at all. The cautionary tales are legion. Diamond
is only metastable; it should, given enough time, turn into graphite, and the
fact that it does not is simply a matter of astronomically slow dynamics between
two equilibrium states. More generally, complex systems exhibit an extremely
wide spectrum of time scales: there is always ``something'' happening, even
without any external shock, and equilibrium is never truly reached as the system
slowly explores local attractors without ever finding the best one --- a
phenomenon known as \emph{weak ergodicity breaking} \cite{bouchaud1992weak}.
Richard Feynman's pithy definition captures the difficulty exactly:
\emph{equilibrium is when all the fast things have happened but the slow things
have not}. Whether the ``fast'' things are really fast, and whether the ``slow''
things matter for economics, are precisely the open questions. Assuming fast
equilibration is not an innocuous simplification: it suppresses history,
path-dependency and coordination failures --- which, as the rest of this review
argues, can in fact be the essential features rather than negligible details.

Indeed, the last fifty years of statistical physics have taught us that large,
complex, interacting systems do \emph{not} generically converge to a simple,
stable equilibrium at all. There is, on the contrary, a whole zoology of generic
behaviours in which the dynamics fundamentally spoils the equilibrium narrative.
It is useful to organize this zoology into three broad mechanisms --- three
``flavours'' of out-of-equilibrium behaviour --- that will recur throughout this
review.

\paragraph{(A) Multiplicity and self-trapping.}
When a system possesses many --- often exponentially many --- competing
equilibria, its dynamics can get trapped. It may converge only after
astronomical times; or it may appear to converge and then, much later, switch
to another attractor, a phenomenon known as \emph{aging}
\cite{bouchaud1998out,cugliandolo1994evidence}. The state that is eventually
selected then depends on the initial conditions and on the entire history of
the system. Glasses, spin glasses and constraint-satisfaction problems are the
paradigmatic physical examples; we shall meet economic cousins in self-trapping
by habit (Sec.~\ref{sec:habit}), in the quasi-nonergodicity of belief dynamics
(Sec.~\ref{sec:prophecies}), in the SK game (Sec.~\ref{sec:skgame}) and in the
rewiring of supply networks (Sec.~\ref{sec:rewiring}).

\paragraph{(B) Non-convergence.}
The dynamics may simply be unable to converge at all. Even when an equilibrium
exists, it may be dynamically inaccessible (for example if it is \emph{linearly
unstable}, but other scenarios exist), the system settling instead into
perpetual oscillations or deterministic chaos. These are genuinely
\emph{endogenous} fluctuations, requiring no external shock whatsoever. This is
the generic fate of large systems of interacting units with non-reciprocal
(asymmetric) couplings --- from random neural networks
\cite{sompolinsky1988chaos,van1996chaos} to complex games
\cite{galla2013complex,sanders2018prevalence,PhysRevX.14.021039} --- and, as we
shall see, of firm networks driven by plausible t\^atonnement dynamics and
rewiring (Sec.~\ref{sec:tatonnement}).

\paragraph{(C) Self-organized criticality.}
Finally, and most subtly, the dynamics may spontaneously drive the system
\emph{towards} a fragile, marginally stable state, poised at the very edge
between stability and instability. This is the self-organized criticality (SOC)
paradigm of Bak and collaborators \cite{bak1987self,bak2013nature}, and its
economic incarnation was anticipated by Bak, Chen, Scheinkman and Woodford
\cite{bak1993aggregate,scheinkman1994self} and, in a different language, by
Minsky \cite{minsky2008stabilizing}, see \cite{bouchaud2024self} for a recent
review. At such critical points, insignificant perturbations can generate large
disruptions through the propagation of ``avalanches'' across the system, and
fluctuations are fat-tailed and long-ranged. We devote Sec.~\ref{sec:soc} to
this mechanism.

These scenarios are not exotic curiosities; they are \emph{generic}. The
intuition, or conjecture, that runs through this review is therefore the
following: understanding and modelling economic and financial swings requires
genuine non-equilibrium features. The burden of proof, I would argue, should
lie with those who claim that a large economy --- a system of millions of
heterogeneous, interacting, adapting agents --- should nonetheless sit quietly
in a unique, stable, rapidly-reached equilibrium.

It is worth having in mind a minimal mathematical picture of why marginal
stability and large fluctuations go hand in hand. Consider the deviation
$x(t)$ of some quantity from its equilibrium value, obeying
$\dot{x} = -\kappa x + \eta$, with $\eta$ a white noise of amplitude $\sigma$.
This Ornstein--Uhlenbeck process has stationary variance
$\mathbb{E}[x^2] = \sigma^2/2\kappa$ and relaxation time $\kappa^{-1}$. As the
system approaches an instability, $\kappa \to 0$, \emph{both} the variance of
the fluctuations \emph{and} the relaxation time diverge, as $\kappa^{-1}$. The
multidimensional generalization, with a stability matrix $\mathbb{K}$ whose
least-stable eigenvalue has real part $-\kappa^\star$, behaves identically as
$\kappa^\star \to 0$. The lesson is universal and will echo through
Secs.~\ref{sec:firms}--\ref{sec:soc}: {near marginal stability, a system
both amplifies exogenous shocks and becomes correlated over very long times.}
Excess volatility and critical slowing-down are two sides of the same coin.

%=====================================================================
\section{Multiple Equilibria, Mimetism, and Self-Trapping}
\label{sec:multiplicity}
%=====================================================================

The simplest way in which the tidy equilibrium picture breaks down is through
the \emph{multiplicity} of equilibria induced by interactions between agents.
When agents care about what other agents do --- through imitation, social
influence, or simple strategic complementarity --- the aggregate can support
several self-consistent states, and which one is realized becomes a matter of
history rather than of fundamentals.

\subsection{Interactions and multiple equilibria: a random-field Ising toy model}
\label{sec:rfim}

Consider a deliberately over-simplified ``restaurant conundrum'' with social
interactions. A continuum of agents must decide whether or not to patronize a
restaurant. Agent $i$ has an idiosyncratic \emph{willingness to pay} $h_i$,
drawn from some distribution, and is further attracted by the popularity of the
place: the utility of the decision to go increases with $\phi \in (0,1)$, the
restaurant's \emph{filling rate}, with a coefficient $J$ measuring the taste for
well-attended places. An agent goes if $h_i > p - J\phi$, where $p$ is the
price. High attendance reduces the effective price of the menu, through
coefficient $J$.

This immediately yields a self-consistent equation for the filling rate,
\begin{equation}
    \phi = \mathbb{P}[\, h > p - J\phi \,],
\end{equation}
which is nothing but the mean-field equation of a random-field Ising model, and
has been rediscovered many times in the discrete-choice-with-social-interactions
literature \cite{brock2001discrete,gordon2009discrete,semeshenko2008collective},
see \cite{bouchaud2013crises} for a review.

The phenomenology is classic but instructive. When $J$ is small (weak social
interaction), there is a single solution $\phi^\star$ for every price $p$. But
when $J$ is large enough, there is a whole range of prices $[p_<,p_>]$ for
which \emph{two} stable equilibria coexist --- a well-attended restaurant and a
nearly empty one. The realized state then depends on history: as the price is
slowly raised and lowered, the filling rate traces out a \emph{hysteresis}
loop \cite{bouchaud2013crises}. The system is, in the language of Sec.~\ref{sec:dynamics}, exhibiting
flavour~(A): multiplicity plus history-dependence. There is no longer a
one-to-one map from fundamentals to outcomes. 

Note that this multiplicity is of a very different nature from the ``sunspot''
indeterminacy familiar from the classical macroeconomic literature
\cite{farmer1999macroeconomics,gali2015monetary}. There, a unique steady state
is locally \emph{indeterminate} because, in the linearized rational-expectations
dynamics, too many eigenvalues lie inside the unit circle: the so-called transversality
(``no-explosion'') condition at infinity then fails to pin down the
forward-looking variables, leaving a continuum of bounded equilibria within which payoff-irrelevant
``sunspot'' signals can move the economy. Much of the literature is
devoted to imposing conditions --- such as the Taylor principle --- that
\emph{restore} determinacy. 

The multiplicity I have in mind here is the opposite: it
is not a fragile local indeterminacy around one steady state, but a genuine
coexistence of \emph{distinct} macroscopic states (a crowded and an empty
restaurant), robust over a whole region of parameter space, and selected by
history through hysteresis. Genericity, not fine-tuning, is the rule --- a theme
to which we return throughout.

\subsection{Confidence collapse may induce crises}
\label{sec:collapse}

The very same type of mechanism may operate at the scale of an entire economy, as had been advocated by, e.g. De Grauwe \cite{de2011animal}. Morelli,
Benzaquen, Tarzia and Bouchaud \cite{morelli2020confidence}
studied a multi-household DSGE-like model in which past aggregate consumption
feeds the \emph{confidence}, and therefore the consumption propensity, of
individual households --- a dose of self-reflexivity injected into an
otherwise orthodox equilibrium model. This suffices to generate a strikingly
rich variety of realistic output dynamics: a regime of high output with no
crises; a regime of high output with excess volatility and deep, short-lived
recessions driven by self-reflexivity; and a genuinely bistable regime in which a relatively mild drop in
economic conditions can tip the economy into a temporary \emph{confidence
collapse} and a steep decline in activity. Crucially, the crisis probability
depends \emph{exponentially} on the parameters of the model --- which means that
markets cannot efficiently price the associated risk premium. (We shall meet
this exponential rarity of endogenous crises again in
Sec.~\ref{sec:prophecies}.) The authors conclude that, in such a world,
\emph{narratives} may become an important monetary-policy tool that can help
steer the economy back on track --- a point I return to in
Sec.~\ref{sec:conclusion}. The static bistability described here is the
equilibrium face of a dynamic learning process; its dynamic counterpart is the
subject of Sec.~\ref{sec:prophecies}, where a generalisation of Kirman's famous
ants model will be discussed \cite{kirman1993ants,moran2020schrodinger}.

\subsection{Cliff-edge optimization and fragility: an appetizer}
\label{sec:appetizer}

Return for a moment to the restaurant of Sec.~\ref{sec:rfim}, but now from the
point of view of the \emph{owner} setting the price. When $J$ is large, the
owner can raise the price all the way up to $p = J$, increasing profit
\emph{without losing clients}, precisely because the popularity of the place
($J$) sustains attendance. The optimal price is therefore right at the edge of
the region of bistability. But this optimum sits on a \emph{cliff-edge} \cite{moran2025critical}: any
small additional price increase above $J$ triggers an \emph{irreversible}
collapse of attendance --- and there are no precursors, no early warning, just
as a brittle material fractures without the gradual yielding of a ductile one.
Optimization has bred fragility. This is a first glimpse of a theme ---
\emph{often, optimization leads to a lack of resilience} --- that we develop
fully in Sec.~\ref{sec:soc}.

\subsection{Self-trapping by force of habit}
\label{sec:habit}

Multiplicity need not come from social interaction; it can come from an agent's
own \emph{history}. Rational-choice theory assumes agents maximize a fixed
utility function, but exploring the universe of available choices is costly, and
in practice memory effects and habit formation may dominate over genuine
optimization. Moran, Fosset, Luzzati, Bouchaud and Benzaquen
\cite{moran2020force} studied a stylized model with a \emph{history-dependent}
utility function, in which the utility of a given choice is increased each time
that choice has been made in the past, with a decaying memory kernel.

The result is that self-reinforcing effects can cause an agent to get stuck with
a choice \emph{by sheer force of habit}: an objectively suboptimal option
becomes a subjective attractor, as memory slowly chisels the utility landscape
into a deep well around whatever the agent happens to have been doing in the past. There is
a sharp transition between a phase of free (``ergodic'') exploration of the space of choices
and a self-trapped, ergodicity broken phase. Remarkably, at the transition the distribution of
trapping times is precisely a Zipf law, and the self-trapped phase exhibits
\emph{super-aging} behaviour. This is a clean micro-level instance of
flavour~(A) --- broken ergodicity, history dependence --- that appears in many
models with self-reinforcement, like in the canonical example of Polya urns (see
\cite{pemantle2007survey} for a review). This mechanism is a direct cousin of
what is at play in the self-fulfilling prophecy model of Sec.~\ref{sec:prophecies} (see also \cite{harris2015random} for a related example). Broken-ergodicity and aging will also appear in Sec.~\ref{sec:skgame} and Sec.~\ref{sec:rewiring}, although the underlying mechanism is different and
induced by the ``radical complexity'' of the situation agents have to face.

%=====================================================================
\section{Learning in a Complex World}
\label{sec:learning}
%=====================================================================

If agents are not naturally endowed with rational expectations, perhaps they can
\emph{learn} them. Indeed, the standard defence of the rational-expectations
framework is dynamical: as Evans and Honkapohja put it, \emph{``in standard
macroeconomic models, rational expectations can emerge in the long run,
provided the agents' environment remains stationary for a sufficiently long
period''} \cite{evans2013learning,marcet1989convergence}. Two words in this
sentence bear all the weight: ``long run'' and ``stationary''. What is ``long run''? Days, weeks, months, years, centuries? And can this be short enough for the world to be``stationary''?  This section is
largely about the second: in a genuinely complex world, are complex situations learnable at all, even if the environment is stationary?\footnote{The fact that, even in classical rational expectation models, economic agents (and the modeller!) face intractably complex mathematical problems, has been recently pointed out by Ben Moll \cite{moll2026trouble}.}   

\subsection{Self-fulfilling prophecies and quasi-nonergodicity}
\label{sec:prophecies}

The difficulty is most acute when the beliefs of agents \emph{influence} the
very outcomes they are trying to predict --- think of business or consumer
confidence, or the confidence of players in a football team. Bouchaud and
Farmer \cite{bouchaud2023self} studied a model of an exchange economy in which $N$
agents trade assets contingent on an observable signal whose probability
depends on public opinion. In its simplest form, the true probability of the
event at time $t+1$ is a function of the average subjective belief,
$\mathbb{P}(t+1) = F\!\big(\sum_i \mathbb{P}_i(t)/N\big)$, where $F$ is a
sigmoidal curve. Each agent updates beliefs in response to observed outcomes,
discounting the distant past at some rate $\lambda$. Agents make occasional
errors, or die and are replaced by newborns with random priors, with probability
$\varepsilon$.

The results overturn the learning defence of rational expectations. The
distribution of the observed signal is described by a \emph{quasi-nonergodic}
process: agents continue to \emph{disagree with each other forever}. At large
times the aggregate belief \emph{appears} to converge to one of the fixed
points of $F$, but ``crises'' --- sudden reversals --- keep happening at a small
rate $\sim \lambda\, e^{-A/\lambda}$, where $A$ is a constant. In the limit
$\lambda \to 0$ the eventual outcome becomes strictly \emph{unknowable}: a
simple, concrete model of \emph{radical uncertainty}
\cite{bouchaud2021radical,king2020radical}. (The special case $F(x)=x$ maps onto
Kirman's celebrated ants model, with its spontaneous flip-flops between the two
states \cite{kirman1993ants,kirman2000learning}). The exponential rarity of these
endogenous crises is exactly the same phenomenon encountered at the macro scale
in the confidence-collapse model of Sec.~\ref{sec:collapse} \cite{morelli2020confidence} ---
the same physics, seen through learning dynamics rather than static bistability.

\subsection{Market failure and wealth inequality}
\label{sec:wealth1}

There is an interesting corollary. If agents disagree forever, then by the same
token they are willing to \emph{trade} with each other forever. In their paper,
Bouchaud and Farmer \cite{bouchaud2023self} show that a market organized so that
agents trade according to their beliefs reveals not the fair, flat average
belief, but the \emph{wealth-weighted} average --- and these are not the same.
Worse, because wealth dynamics is \emph{multiplicative}, successful bold bets
are highly profitable, and some agents temporarily become extremely rich while
others become extremely poor. Disagreement thus spontaneously generates a
\emph{power-law distribution of wealth}, which makes the wealth-weighted average
significantly different from the flat average. The market fails in a precise
sense: prices do not reveal the true probability, and do not help agents
coordinate on ``truth''.

This is one of several routes by which out-of-equilibrium dynamics generates
inequality endogenously; we shall meet another, in a primitive production
economy, in Sec.~\ref{sec:inequality}. Both are ultimately rooted in the same
multiplicative-growth condensation mechanism identified long ago by U. Yule,
H. Simon, H. Kesten and others (see e.g. \cite{bouchaud2000wealth}). Inequality
here is not an input assumption but an \emph{emergent symptom} of market failure.

\subsection{The SK game: unlearnable games and satisficing decisions}
\label{sec:skgame}

How hard is learning, really, in a complex world? To answer sharply,
Garnier-Brun, Benzaquen and Bouchaud \cite{PhysRevX.14.021039} introduced the
\emph{Sherrington--Kirkpatrick game}, a discrete-time binary-choice model
inspired directly by mean-field spin glasses. The ambition is to propose ``a simple model of a complex world''. There are $N$ interacting agents,
each choosing a strategy $S_i = \pm 1$; the reward of agent $i$ is
$S_i \sum_j J_{ij} S_j$, where the influence matrix $J_{ij}$ encodes the
structure of the game. When $J_{ij}=J_{ji}$ the interactions are
\emph{reciprocal} --- cooperative if positive, non-cooperative if negative;
when $J_{ij}=-J_{ji}$ they are \emph{anti-reciprocal}, i.e.\ competitive, of a
predator--prey type. Agents learn via Sato--Crutchfield reinforcement-learning
dynamics \cite{sato2005stability} for a ``propensity'' $Q_i$ to play $+1$, with
$m_i := \langle S_i \rangle = \tanh(\beta Q_i)$; for small learning rates this
reduces to {\it naive mean-field} equations
$m_i^{t+1} = \tanh\!\big(\beta \sum_j J_{ij} m_j^t\big)$, which are exact in this
case.

The central result runs counter to the rational-expectations hope. Even in a
\emph{completely static} environment, agents are unable to learn collectively
optimal strategies. This is the curse of ``radical complexity''
\cite{bouchaud2021radical,sharma2021good,colon}, that we will meet again in
Sec.~\ref{sec:rewiring}. Either the learning process gets trapped in a suboptimal
fixed point, or it never converges and leads to a never-ending evolution of
intentions. For random (spin-glass) couplings and reciprocal interactions,
above a critical $\beta_c$ there is an \emph{exponential} number of Nash-like
equilibria $\sim e^{\Sigma(\beta) N}$; the states actually reached by learning
are suboptimal, fragile and --- because they depend chaotically on initial
conditions and on the $J_{ij}$ --- effectively \emph{unknowable}. Only a
centralized, omniscient planner endowed with formidable computing power (a
Parisi ground-state solver \cite{mezard1987spin}) could determine the optimum.
The full phase diagram is rich: as competition (non-reciprocity) is increased,
stable fixed points give way first to chaos and, in the high-competition limit,
to quasi-cycles; there is also a ``paramagnetic'' phase converging to the trivial
rock-paper-scissors equilibrium $m=0$, and, over a large swath of this static
parameter space, spontaneous \emph{aging}, i.e. a permanent state of non-equilibrium probing deeper and deeper ``valleys'' in phase space.

Several of the findings are worth stating plainly, because they carry economic
morals. Long memory of past rewards is \emph{beneficial} to learning, whereas
over-reaction to the recent past is \emph{detrimental} and drives the system
into cycles or chaos --- an echo we will hear again in the t\^atonnement
dynamics of Sec.~\ref{sec:tatonnement}. Some amount of randomness in the
learning process, perhaps paradoxically, \emph{improves} the collective
outcome. And hypersensitivity to the game's parameters makes it impossible to
predict \emph{ex ante} who will end up better or worse off. The conclusion is
Herbert Simon's \cite{simon1955behavioral}:
complex situations are generically \emph{unlearnable}, and agents must make do
with \emph{satisficing}, rather than optimal, solutions. The SK game is, in a
sense, the conceptual capstone of this review: a single, static model that
simultaneously displays all three flavours of Sec.~\ref{sec:dynamics} ---
trapping and aging (A), chaos and quasi-cycles (B), and the marginal stability
(C) that we now pursue in earnest.

%=====================================================================
\section{Coordination Breakdown in Firm Networks}
\label{sec:firms}
%=====================================================================

We now turn from abstract games to the concrete architecture of a modern
economy: the network of firms that buy inputs from, and sell outputs to, one
another. Here one can compare, in the same setting, the equilibrium benchmark of
mainstream economics with its out-of-equilibrium alternatives --- and watch the
equilibrium picture dismantle itself in three successive steps.

\subsection{The equilibrium benchmark}
\label{sec:benchmark}

The natural starting point is the Long--Plosser real-business-cycle model,
elaborated into a network setting by Acemoglu, Carvalho and collaborators
\cite{long1983real,acemoglu2012network,carvalho,acemoglu_azar_2020}. A
representative household works and consumes goods $i=1,\dots,N$; firm $i$
produces using labour and the outputs of other firms $j$, through a
Cobb--Douglas production function, defined as
\begin{equation}
\pi_i= z_i \prod_j (Q_{ij}/J_{ij})^{a_{ij}},
\end{equation}
where $\pi_i$ is the production of firm $i$, $z_i$ is the firm productivity,
$Q_{ij}$ is the number of goods firm $i$ buys from firm $j$ and $J_{ij}, a_{ij}
\geq 0$ are weight parameters.

Households maximize utility, firms maximize profits, and all goods and labour
markets clear. The upshot is a well-defined equilibrium for \emph{any} network
weights $J, a$ and \emph{any} productivities $z$, with strictly positive prices
and productions. Fluctuations in this economy are entirely driven by exogenous
productivity shocks $z(t)$, under the assumption that equilibrium is reached
instantaneously. The production network itself is always stable, and the economy
is a succession of equilibria all perfectly determined by the input-output
network and the individual firm productivities. In the absence of a global shock,
GDP fluctuations decrease much too quickly (as $N^{-1/2}$) to explain the ``small
shocks, large business cycle'' puzzle, except for very unbalanced (``granular''
or ``star-like'') graphs, where hub firms fail to average out and idiosyncratic
shocks survive at the aggregate level
\cite{gabaix2011granular,acemoglu2012network,moran2024revisiting}.

\subsection{CES, feasibility failure, and May's instability}
\label{sec:ces}

The Cobb--Douglas assumption is far more special than it looks. It corresponds
to a high degree of substitutability between input goods, and it is \emph{this}
that guarantees a feasible equilibrium for any network and any productivities.
As soon as one moves to the more general Constant Elasticity of Substitution
(CES) family,\footnote{The Constant Elasticity of Substitution (CES) family of
production functions is defined as
\[
\pi_i = z_i \left(\sum_j a_{ij} \left(\frac{J_{ij}}{Q_{ij}}\right)^{\frac{1}{q}}\right)^{-q} \quad \mbox{with} \quad \sum_{j} a_{ij}=1,
\]
where parameter $q$ captures the substitutability of input goods. The
Cobb-Douglas function corresponds to $q \to \infty$. When $q \to 0^+$, on the
other hand, no substitutes are available and the production function reduces to
the classical Leontief production function:
$\pi_i = z_i \min_{j} \left({Q_{ij}}/{J_{ij}}\right)$.} with lower substitutability
--- down to the Leontief limit, where inputs are strictly complementary and no
substitution is possible --- a feasible competitive equilibrium \emph{ceases to
be guaranteed}. The market-clearing equations may have no solution with all
prices and productions positive; some firms, for all practical purposes, must go
bankrupt. Mathematically, feasibility requires a certain matrix $\mathbb{M}$ to
be an ``M-matrix'', i.e.\ to have all its eigenvalues with non-negative real
parts --- a condition first identified by Hawkins and Simon \cite{hawkins1949note}.

This is exactly the condition studied by Moran and Bouchaud in \emph{``May's
instability in large economies''} \cite{moran2019may}, and it forges an intriguing
link between this section and Sec.~\ref{sec:soc}. Building on Robert May's
original argument for large ecosystems \cite{may1972will}, they show --- using
random-matrix theory, and exploiting the fact that the equilibrium equations are
formally identical to those of a Generalized Lotka--Volterra ecology
\cite{bunin2017ecological,biroli2018marginally,stone2018feasibility} --- that
such firm networks \emph{generically become dysfunctional} as their size
increases, as the heterogeneity between firms grows, or as the substitutability
of inputs is reduced. At marginal stability, the distribution of firm sizes
develops a power-law tail, as observed empirically; and small idiosyncratic
shocks trigger \emph{avalanches} of defaults with a power-law distribution of
total output losses. This scenario is a natural candidate to explain the ``small
shocks, large business cycles'' puzzle --- as anticipated long ago by Bak, Chen,
Scheinkman and Woodford. A conjecture, to which we return in Sec.~\ref{sec:soc},
is that \emph{evolutionary and behavioural forces conspire to drive the economy
towards marginal stability}.

\subsection{A naive dynamical model: endogenous cycles and chaos}
\label{sec:tatonnement}

Postulating that the economy is always in equilibrium is only justified if
convergence towards equilibrium is \emph{fast} compared to the speed at which
the fundamentals (productivity, network, tastes) themselves change. So one must
ask: how fast do imbalances --- of supply and demand, of profit --- actually
wash away? Dessertaine, Moran, Benzaquen and Bouchaud \cite{dessertaine2022out}
endowed the classical firm-network model with plausible, generic dynamical
(``t\^atonnement'') rules by which prices and productions respond to
supply/demand and profit imbalances, and asked under what conditions the network
can \emph{dynamically} attain the competitive equilibrium (see also the discussion in  \cite{mandel2026shock}).

The answer depends on where the system lies in parameter space. For example, the
time needed to reach equilibrium \emph{diverges to infinity} as the system
approaches the very instability point at which the Hawkins--Simon condition is
violated \cite{hawkins1948some,hawkins1949note} --- the same boundary as in
Sec.~\ref{sec:ces}. Such critical slowing-down is itself a source of excess
volatility, through the accumulation and amplification of exogenous shocks
(recall the $\kappa^\star \to 0$ argument of Sec.~\ref{sec:dynamics}).

But there is more. Once essential physical constraints absent from the naive model
--- causality, inventory management, the impossibility of consuming more than is
produced --- are properly accounted for, the dynamically consistent model
displays a rich phase diagram. Depending on the strength of the firms' reaction
to imbalances and on the perishability of goods, the economy can: (a) collapse
entirely; (b) converge to the competitive equilibrium, after a time that may
become very long; or (c) enter a state of \emph{perpetual disequilibrium}, with
purely endogenous fluctuations that may be periodic or fully chaotic ---
reminiscent of real business cycles, and of the coordination breakdown among
firms familiar from managerial ``beer-game'' experiments \cite{Sterman1989}. (A
sliver of deflationary equilibrium also appears.) In all regimes, including
regime (c), {\it the very same standard economic equilibrium prevails} but is
{dynamically out of reach}. This is flavour~(B) in its purest economic form, and
it offers an alternative resolution of the ``small shocks, large business cycle''
paradox: the economy may be permanently in a turbulent state, far from
equilibrium, without necessarily being close to a critical point. Even close to
equilibrium, incidentally, the evolution is only \emph{cone-wise} linear, because
of the constraints --- a reminder that linearization can be treacherous
\cite{dessertaine2022non,martin2026resilient}. Small perturbations can still
generate highly non-linear effects!

It is worth noting \emph{en passant} that the same qualitative picture ---
spontaneous booms and busts, with no exogenous trigger --- emerges in the
stylized macroeconomic agent-based model ``Mark0'' of Gualdi, Tarzia, Zamponi
and Bouchaud \cite{gualdi2015tipping,gualdi2015endogenous}, and in its
constraint-satisfaction cousin \cite{sharma2021good}. There, firms default when
their debt-to-sales ratio exceeds a threshold, the default of one firm
deteriorates the environment of others, and above a critical coupling the
feedback synchronizes bankruptcies into full-blown, endogenous crisis waves,
with spikes of unemployment --- see also \cite{pangallo2025synchronization}. This is a concrete demonstration, within a
fully-fledged (if stylized) ABM, that a market economy can generate its own
business cycle from within.

\subsection{Rewiring: exponentially many path-dependent equilibria}
\label{sec:rewiring}

The models above take the network as given. But a crucial ingredient is
missing: firms fail, new firms are born, and --- most importantly --- firms
\emph{rewire}, choosing their suppliers so as to reduce costs. Colon and
Bouchaud \cite{colon,colon2022radical} asked whether the \emph{topology} of the
production network can itself reach equilibrium under such cost-driven rewiring.
In the textbook Cobb--Douglas economy with constant returns to scale and
unlimited rewiring capacities, Acemoglu and Azar \cite{acemoglu_azar_2020}
had proved that a unique optimal network exists --- a downhill, ``ratchet''
dynamics converging to the best of all worlds. Colon and Bouchaud introduce
three \emph{mild} departures from this idealized setup --- capping the number of
simultaneous supplier switches, allowing deviations from constant returns to
scale, and imposing limited network visibility --- and show that \emph{each one
alone is sufficient to break the result}.

The economy then admits an \emph{exponential} number of path-dependent
equilibria, and shocks trigger \emph{irreversible} reconfiguration waves. The
dynamics is the typical hyper-slow, glassy dynamics of complex systems, with
strong dependence on initial conditions and history, like in the SK-game of
Sec.~\ref{sec:skgame}. Even when the dynamics does converge --- possibly only
after very long times --- the topology it settles on cannot be anticipated from
the initial condition: it depends chaotically on the microscopic details of the
rewiring process, such as the very order in which firms are allowed to update
their suppliers. And the network so selected is itself \emph{fragile}: the
default of a single firm can set off a fresh cascade of rewirings that reshapes
the entire structure. This is flavour~(A) at the level of network topology, driven by the problem complexity: where
Sec.~\ref{sec:tatonnement} showed that a \emph{unique} equilibrium can be
dynamically inaccessible, this subsection shows that once the special
Cobb--Douglas-plus-CRS-plus-full-visibility assumptions are relaxed, the
equilibrium \emph{set itself} becomes exponentially degenerate and history-,
even accident-, dependent. One should thus be wary of general-equilibrium policy
inferences in ``radically complex'' situations.

\subsection{Emergent inequality in a good-exchange economy}
\label{sec:inequality}

As a final variation, Patil and Bouchaud \cite{patil2025emergent} studied a
primitive agent-based good-exchange model in which each agent simultaneously
produces and consumes, with a utility that saturates fast at large consumption,
and in which prices are updated whenever supply fails to match demand. The
system generically exhibits a non-trivial phase transition beyond which a
market-clearing equilibrium \emph{exists but becomes dynamically unreachable}
--- the same ``existence without accessibility'' motif as in
Secs.~\ref{sec:tatonnement}--\ref{sec:rewiring}. When production capacity
exceeds a threshold and adapts too slowly, some agents cannot sell all their
goods; this induces global price \emph{deflation} and drives strong wealth
inequalities, with the spontaneous separation of the population into a rich
class and a poor class. Together with the belief-market inequality of
Sec.~\ref{sec:wealth1}, this closes a loop: heterogeneity plus market feedback
generically produces \emph{wealth condensation}, whether through multiplicative
bets or through slow-adapting production --- inequality as an emergent,
dynamical symptom of out-of-equilibrium market failure, rather than a
distributional assumption.

%=====================================================================
\section{Fragile Optimization and Self-Organized Criticality}
\label{sec:soc}
%=====================================================================

We now develop in full the theme first glimpsed in Sec.~\ref{sec:appetizer}:
that optimization itself, far from delivering a benign equilibrium, often drives
a complex system to a fragile, critical point where it is maximally exposed to
disruption. This is the self-organized criticality paradigm
\cite{bak1993aggregate, bak1996nature, bouchaud2024self}, and it supplies the deepest and most policy-relevant of
our three mechanisms.

\paragraph{What is special about a critical point.}
Recall the branching, or avalanche, process --- a very large class of models
describing sandpile avalanches, epidemics, neuronal cascades, and default or
bankruptcy waves alike. A single ``active'' unit triggers on average $R_0$
others, each of which may trigger more. When $R_0 < 1$, activity dies out and
avalanches are small; when $R_0 > 1$, a single seed can ignite a system-wide
conflagration. Exactly at $R_0 = 1$ --- the critical point --- the distribution
of avalanche sizes is a scale-free power law,
$P(S) \propto S^{-3/2}\exp(-\varepsilon^2 S)$ with $\varepsilon = 1 - R_0$, so
that most avalanches are small but \emph{some} are enormous
\cite{harris1963theory}. The system \emph{looks} stable but occasionally goes
haywire with no apparent cause --- precisely the phenomenology of excess
volatility.

\paragraph{When is criticality not a fine-tuning coincidence.}
The obvious objection is that $R_0=1$ is one special value among a continuum,
and that generically a system is either subcritical (dull) or supercritical
(explosive). The seminal insight of Bak, Tang and Wiesenfeld
\cite{bak1987self} was that the control parameter itself can become a
\emph{dynamical variable}, driven by feedback towards the critical value. A pile
of sand fed slowly from above steepens until its slope reaches the critical
angle, at which avalanches of all sizes maintain it there. An epidemic's $R_0$
is pushed down by containment when cases surge and drifts back up through
complacency when they subside, hovering around unity \cite{sornette1994sweeping}.
In economics and finance the analogue is compelling: \emph{efficiency-seeking
behaviour continually erodes the safety margins that keep the system stable}, as
Minsky anticipated long ago \cite{minsky2008stabilizing,minsky}.

\paragraph{The resilience--efficiency trade-off, concretely.}
Martin, Moran, Panja and Bouchaud \cite{martin2026resilient}
make this precise in a supply-chain network in which firms use non-substitutable
(Leontief) inputs, hold precautionary inventories $\kappa$, face idiosyncratic
productivity shocks of amplitude $\sigma$, and adjust through \emph{quantities}
rather than prices. They show, analytically and numerically, that there is a
\emph{critical boundary} in the $(\sigma,\kappa)$ plane: above the threshold,
the economy absorbs shocks and fluctuates mildly; below it, cascading shortages
make system-wide crises inevitable. Close to the threshold, aggregate output
volatility \emph{diverges} through network-mediated amplification of purely
idiosyncratic shocks --- another concrete mechanism for ``small shocks, large
business cycles''. The crux is the last step: because inventories are
\emph{costly}, competitive pressure continually drives firms to reduce them,
pushing the whole system \emph{towards} the fragility boundary. When $\kappa$ is
above but close to the critical $\kappa_c$, crises are still exponentially rare,
and firms are tempted to trim their insurance still further --- a Minsky
``complacency'' cycle. A resilience--efficiency trade-off emerges, putting the
celebrated gains from lean, ``just-in-time'' supply chains at risk.
(Reassuringly, sufficiently abundant \emph{supplier diversification} shifts the
threshold and can eliminate the fragile regime entirely --- a hint that the
fragility is a design choice, not a fatality, see also the discussion in
\cite{elliott2022networks}).

\paragraph{Timeliness criticality.}
The same logic governs the propagation of \emph{delays} (in e.g. train arrival
times) rather than shortages. Moran et al. \cite{Moran2024,moran2025critical} introduced a minimal model
of delay propagation in a network of tasks, in which the delay at a node is
\begin{equation}
    \tau_i(n{+}1) = \big[\max_{j\in\partial_i}\tau_j(n) - B\big]^+ + \epsilon_i,
\end{equation}
with $B$ a time buffer and $\epsilon_i$ are idiosyncratic independent positive
delay shocks and $\partial_i$ the set of nodes on which node $i$ relies. 

Above a critical buffer $B_c$, delays self-heal; below it, they
accumulate without bound, and close to $B_c$ delay ``avalanches'' of all sizes
appear. As with inventories, competitive and efficiency pressures push operators
to shave buffers --- train timetables, airline schedules, bank capital cushions
--- until a major breakdown occurs, regulators step in, and the cycle repeats.
The mechanism is generic across socio-technical systems \cite{dekker2021cascading}.

\paragraph{The many faces of the same idea.}
This scenario recurs across the models of this review. The CES feasibility
boundary of Sec.~\ref{sec:ces} and the instability of supply-chains when
precautionary inventories are too scarce are in spirit the \emph{same} critical point,
reached because a growing, complexifying, profit-seeking economy generically
erodes its own stability margin --- a new firm added at constant productivity can
only \emph{decrease} the least stable eigenvalue, so that a growing economy
becomes more unstable unless productivity rises fast enough. The fragility of
financial-market liquidity has the same flavour: competition drives
market-makers' spreads towards break-even, making them hypersensitive to any
uptick in volatility, so that \emph{more volatility begets less liquidity begets
more volatility} \cite{wyart2008relation,fosset2020endogenous,bouchaud2018trades}.
Inflation, too, can proceed in \emph{avalanches}, when the repricing of one firm
pushes its customers towards repricing, with a branching ratio that empirical
calibration places startlingly close to unity
\cite{nirei2024repricing,leal2021repricing}. And systems governed by antagonistic
excitatory and inhibitory forces --- from the balancing of a stick
\cite{patzelt2011criticality,cabrera2004human} to neuronal activity
\cite{lombardi2017balance,chialvo2010emergent} --- generically operate near
criticality: \emph{the better one stabilizes such a system, the more difficult it
becomes to predict, generating occasional large excursions}.

The general moral, made forcefully by Carlson and Doyle in their ``Highly
Optimized Tolerance'' framework \cite{carlson1999highly} and echoed by Taleb's
notion of anti-fragility \cite{taleb,hynes2022systemic}, is that {the quest for
efficiency and the necessity of resilience may be mutually incompatible.}
Optimization, in a complex system, tends to consume the very redundancy that
would protect it. Fragility and efficiency are two sides of the same coin.

%=====================================================================
\section{Discussion and Conclusion}
\label{sec:conclusion}
%=====================================================================

\emph{Is the concept of ``equilibrium'' relevant in complex economies?} From
analogies with physical complex systems, and from the several stylized models
reviewed here, one is tempted to answer: not much. Generic dynamics does not, in
general, lead to a unique, stable, rapidly-reached equilibrium. Convergence to
equilibrium appears to be the exception rather than the rule. Excess volatility,
endogenous crises and crashes, and inflation swells are, in this light, most
naturally understood as \emph{non-equilibrium phenomena}. Blanchard's ``dark
corners'' \cite{blanchard2014danger}, I would suggest, are not rare pathologies
to be visited only in extremis; they may be where a complex economy generically
lives.

\paragraph{Three routes, three flavours.}
It is worth summarizing how the models fall into the taxonomy of
Sec.~\ref{sec:dynamics}, because the classification is not merely tidy --- it
has some empirical consequences.

\begin{table}[h!]
\centering
\small
\begin{tabularx}{\textwidth}{@{}lXcX@{}}
\toprule
\textbf{Model} & \textbf{Mechanism} & \textbf{flavour} & \textbf{Fingerprint} \\
\midrule
\midrule
RFIM restaurant / DSGE confidence (\S\ref{sec:rfim})
 & mimetism, multiplicity & A & hysteresis, bistable collapse, exponentially rare crises \\
 \midrule
 
Self-trapping by habit (\S\ref{sec:habit})
 & history-dependent utility & A & ergodicity breaking, history dependence \\
  \midrule
Self-fulfilling prophecies (\S\ref{sec:prophecies})
 & reflexive learning & A+B & quasi-nonergodicity, rare reversals, wealth inequalities \\
  \midrule
SK game (\S\ref{sec:skgame})
 & unlearnable game, radical complexity & A+B+C & exp-many fragile states, chaos, aging \\
  \midrule
T\^atonnement firm network (\S\ref{sec:tatonnement})
 & dynamical inaccessibility & B & endogenous cycles/chaos \\
  \midrule
Good-exchange economy (\S\ref{sec:inequality})
 & market-clearing failure & B & dynamically unreachable equilibrium, wealth inequalities \\
  \midrule
Rewiring network (\S\ref{sec:rewiring})
 & topological multiplicity, radical complexity & A & history-dependence, glassy dynamics \\
  \midrule
May / CES / inventories (\S\ref{sec:ces},\,\S\ref{sec:soc})
 & marginal stability, SOC & C & default avalanches, power-laws, long memory \\
\bottomrule
\bottomrule
\end{tabularx}
\end{table}

There are (at least) \emph{two} distinct out-of-equilibrium routes to excess
volatility, and they make \emph{different} predictions. The first is purely
endogenous cycles and chaos (flavour~B), arising from over-reaction, coordination
breakdown and non-linearity; when a system evolves chaotically far from any
critical point, one does \emph{not} necessarily expect power-law statistics or
long-memory correlations. The second is self-organized criticality (flavour~C);
here fluctuations \emph{should} be fat-tailed and correlated over long time
scales, corresponding to the propagation of avalanches over large portions of
the system. Distinguishing the two is a matter for data, not for theory alone.

\paragraph{The coordination trap.}
There is a subtler, and in some ways more troubling, consequence of collective
instability. In a stable equilibrium world, an agent's local, seemingly
innocuous decision --- a firm shaving its inventory, cutting a supplier, nudging
a price --- has local, proportionate consequences. But in the presence of the
collective instabilities reviewed here, this is no longer true: a small local
decision can bubble up, through network propagation and feedback, into something
catastrophic at the aggregate level. The t\^atonnement dynamics of
Sec.~\ref{sec:tatonnement}, the good-exchange economy of Sec.~\ref{sec:inequality},
the rewiring dynamics of Sec.~\ref{sec:rewiring} and the precautionary
inventories story of Sec.~\ref{sec:soc} are all cases in point --- the individual
has no way of foreseeing that their myopically reasonable choice, aggregated with
everyone else's, will tip the system into turbulence, deflation or collapse.
Coordination is precisely what is needed for a good equilibrium to be reached;
but the flip side is that \emph{miscoordination} --- which no single agent can
detect from their own vantage point --- can unwittingly lead to havoc, like in the famous Schelling model of segregation \cite{grauwin2009competition}.

The rewiring problem of Sec.~\ref{sec:rewiring} sharpens the point into a
dichotomy. In the idealized Cobb--Douglas world of Acemoglu and Azar,
coordination is \emph{easy}: there is a unique optimal network, and cost-driven
rewiring marches downhill towards it like a ball rolling to the bottom of a bowl.
But this is a knife-edge, ``corner'' case. Under the mildest realistic departures
--- capped switching, non-constant returns, limited visibility --- the landscape
shatters into an \emph{exponential} number of Nash equilibria, and there is no
longer any way for decentralized agents to coordinate on the optimal one, any
more than a glass can find its crystalline ground state. The optimum continues to
exist; it has simply become \emph{unknowable} and \emph{unreachable} in practice,
attainable only by the fictitious omniscient planner we already met in the SK game
(Sec.~\ref{sec:skgame}). Genericity, once again, trumps the corner theorem: the
easy-coordination case is the exception, not the rule. 

\paragraph{Toward empirical discrimination --- and the absence of a smoking gun.}
I want to be candid: at this juncture we have many more \emph{narratives} than
we have hard empirical evidence. The models reviewed here are toy models ---
phenomenological scenarios that show excess volatility, crises and inequality
\emph{can} arise naturally out of equilibrium, not calibrated theories that
prove they \emph{do}. This is, I would argue, a respectable and even necessary
stage in the physics of a new problem: one first establishes what is generic and
possible, then confronts it with data. The fingerprints to look for are
reasonably clear --- the statistics of price jumps \emph{not} associated with
news \cite{joulin2008,marcaccioli2022exogenous,cutler1988moves}, the Omori-like
relaxation and near-critical feedback measured through ARCH and Hawkes
calibrations, where the endogeneity parameter is consistently found close to
unity \cite{hardiman2013critical,filimonov2012quantifying,Wehrli2022}, or the
repricing-avalanche calibrations that place the inflation branching ratio at
$R_0 \approx 0.9$ \cite{leal2021repricing}. A renewed, data-driven effort --- in
the spirit of the endo/exo classification of jumps and of the repricing-avalanche
work --- is needed to convince the economics community that \emph{fragility},
rather than a procession of large invisible shocks, is what drives excess
fluctuations. The burden of proof, I maintain, cuts both ways empirically; but
the equilibrium camp has the harder case to make.

I have deliberately kept financial markets in the background of this review,
focusing on the real economy and on generic collective mechanisms. But it is in
financial markets that the SOC scenario finds its most developed empirical
incarnation --- through critical liquidity provision, self-exciting Hawkes
feedback, market-ecology models and contagion through overlapping portfolios. I
refer the reader to a companion review for a fuller discussion
\cite{bouchaud2024self}, and note only that the excess-volatility puzzle in
finance and the small-shocks puzzle in macroeconomics are, in my view, two facets
of a single phenomenon.

\paragraph{Policy.}
If economies and markets are indeed fragile --- poised near critical points by
the relentless pressure for efficiency --- then the policy implications are
significant, see \cite{elliott2022networks}. Any welfare or objective function
that operators, regulators and policymakers seek to optimize should contain an
explicit measure of \emph{robustness} to small perturbations, to parameter
uncertainty, and to tail events. Such a resilience penalty will, for sure,
increase costs and degrade strict economic performance in normal times --- but it
will keep the system a safe distance from the cliff-edge \cite{moran2025critical}.
The goal, in Taleb's language, should be \emph{anti-fragile} systems. There are
gentler levers too: as the confidence-collapse model of Sec.~\ref{sec:rfim}
suggests, in a world of self-fulfilling expectations \emph{narratives} themselves
may become a monetary-policy instrument; and the supplier-diversification result
of Sec.~\ref{sec:soc} shows that fragility can sometimes be engineered away by
design. Even a modest ``tax'' that helps agents coordinate can, in models rife
with bad equilibria, avert the worst outcomes \cite{grauwin2009competition}.

\paragraph{A paradigm shift.}
Ultimately, what is at stake is a shift in modelling philosophy --- from the
elegant ``corner theorems'' of general-equilibrium theory, which establish the
existence and optimality of equilibrium under stringent assumptions, towards an
approach in which \emph{genericity trumps corner theorems}. Large, complex,
adaptive systems do not naturally sit in optimal equilibria; they wander,
oscillate, age, and occasionally crash. Economics, I believe, has much to gain
by taking this seriously --- by treating the economy, in Doyne Farmer's phrase
\cite{farmer2024making}, as a system whose chaos we must learn to make sense of,
rather than as a machine that reliably returns to an optimal state of well-being. \emph{Eppur si muove}.

%=====================================================================
\subsection*{Acknowledgements}
%=====================================================================
I want to thank many collaborators on these topics, in particular M. Benzaquen,
C. Colon, Th. Dessertaine, R. Farmer, A. Fosset, J. Garnier-Brun,
S. Gualdi, D. Martin, J. Moran, F. Morelli, D. Panja, N. Patil, F. Pijpers,
M. Tarzia, U. Weitzel, F. Zamponi. Very useful discussions, comments and
feedback from the participants of the two-day conference on ``Persistent Puzzles
and Paradoxes in Economics'', held at CFM at the end of June 2026 and where this
work was presented, are also warmly acknowledged --- in particular L. Bocquet, V. Carvalho, S. Chelly, 
J. D. Farmer, X. Gabaix, F. Geerolf, R. Koijen, D. Luongo, P. Michailliat, B. Moll, M. Pangallo, L. Pedersen, L. Straub and D. Thesmar. I also thank Rosemary Harris for inviting me to write this short perspective for Europhysics Letters.

\bibliographystyle{unsrt}
\bibliography{references}

\end{document}